\pdfoutput=1
\documentclass[preprint,5p]{elsarticle}
\usepackage{graphicx}
\usepackage{hyperref}
\usepackage{amsmath,amsfonts}
\usepackage{flushend}
\usepackage{color}

\begin{document}

\newcommand{\beq}{\begin{eqnarray}}
\newcommand{\eeq}{\end{eqnarray}}
\newcommand{\non}{\nonumber\\}
\newcommand{\D}{\mathcal{D}}
\newcommand{\p}{\partial}
\newcommand{\Tr}{{\rm Tr}}
\newcommand{\diag}{{\rm diag}}
\newcommand{\sign}{{\rm sign}}
\newcommand{\sech}{{\rm sech}}

\title{Effective field theories on solitons of generic shapes}
\tnotetext[t1]{\tt NORDITA-2014-86}

\author{Sven Bjarke Gudnason${}^{1,2}$}
\ead{bjarke(at)impcas.ac.cn}
\author{Muneto Nitta${}^3$}
\ead{nitta(at)phys-h.keio.ac.jp}
\address{${}^1$Institute of Modern Physics, Chinese Academy of Sciences,
  Lanzhou 730000, China}
\address{${}^2$Nordita, KTH Royal Institute of Technology and
    Stockholm University, Roslagstullsbacken 23, SE-106 91 Stockholm,
    Sweden}
\address{${}^3$Department of Physics, and Research and
    Education Center for Natural Sciences, Keio University, Hiyoshi
    4-1-1, Yokohama, Kanagawa 223-8521, Japan}

\begin{abstract}
A class of effective field theories for moduli or collective
coordinates on solitons of generic shapes is constructed. 
As an illustration, we consider effective field theories living on
solitons in the O(4) non-linear sigma model with higher-derivative
terms. 
\end{abstract}


\maketitle


\section{Introduction}

Effective field theory is one of the most useful tools available to
date. Even the standard model, although renormalizable in its present
formulation, may also be just an effective theory of Nature where
possible supersymmetric and/or grand unified extensions have been
integrated out.
For particles of accessible energies, we can neglect gravity and
consider particles on flat space as a(n extremely) good
approximation. 
This is just a consequence of the separation of scales between the
particle mass and energy versus the scale of gravity, i.e.~the Planck
mass. 
Light fields do not only exist in all of spacetime but  
are sometimes confined to certain subspaces.
For solitons hosting moduli, there is again a situation where
separation of scales can be exploited; namely the mass of the soliton
versus massless or light moduli. 
Effective field theories for moduli have been constructed for many
kinds of solitons, but very often only in cases where the soliton has
a simple, flat or straight shape. 
As examples, the effective actions for monopole moduli
\cite{monopoles}, domain-wall moduli 
\cite{Chibisov:1997rc,DWs,Eto:2006pg,Eto:2006uw}
and for orientational moduli of non-Abelian strings 
\cite{NAstrings,Eto:2006pg,Eto:2006uw,Fujimori:2010fk} have
been constructed. 
When solitons are particle-like such as monopoles
this can describe the low-energy dynamics of the solitons in a compact way 
as geodesics of moduli spaces \cite{monopoles},
while for solitons being extended objects such as domain walls or vortices,
this describes field theories on their world-volume, 
as in the case of D-branes in string theory or more general branes. 
Solitons can, however, generically possess much more complicated
shapes.

In this Letter we construct a first attempt of effective field
theories in principle applicable to solitons of generic shapes and
apply it to a class of models possessing soliton solutions of flat,
spherical, cylindrical and toroidal shapes.

\section{General considerations}

\begin{figure}[!htb]
\begin{center}
\includegraphics[width=\linewidth]{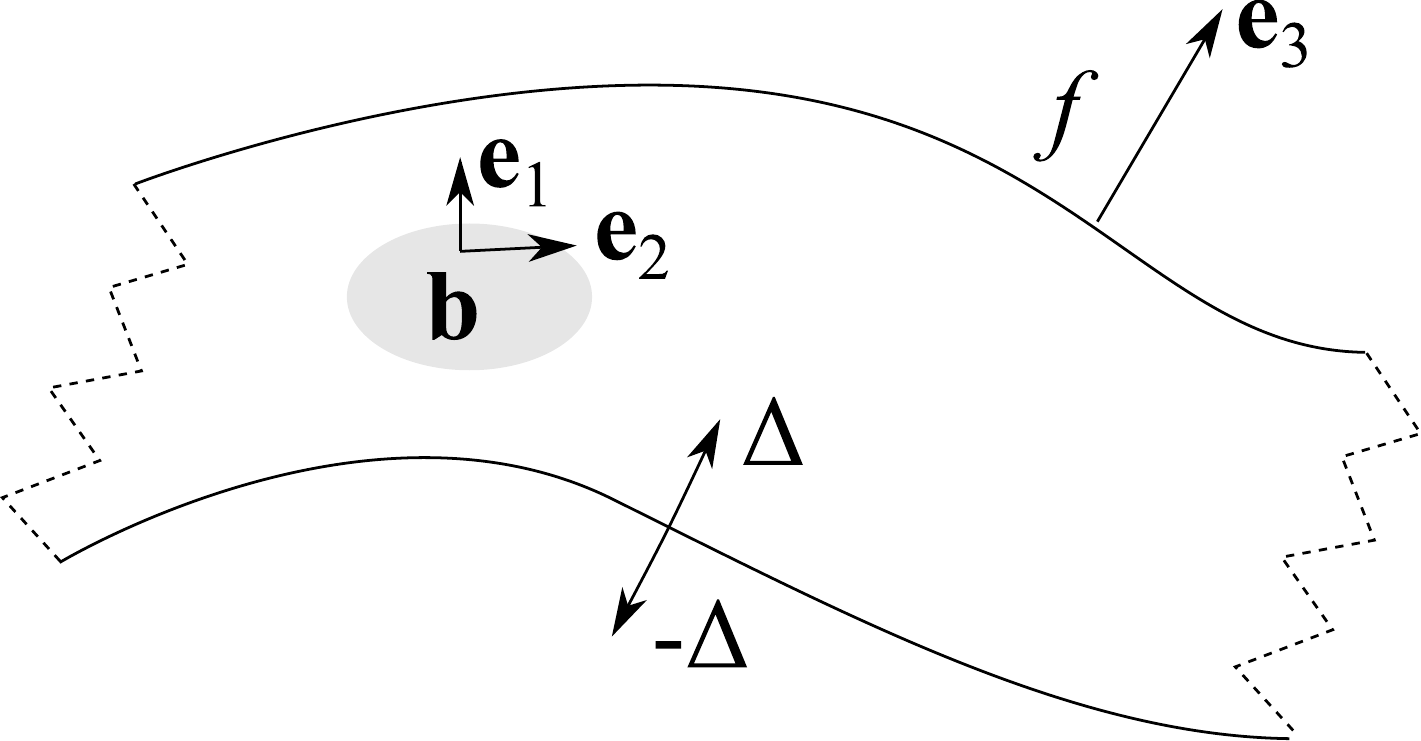}
\caption{Sketch of a generic soliton profile in direction
  $\mathbf{e}_3$. The moduli $\mathbf{b}$ live on the manifold spanned
  by the host soliton. The integration over the codimension is done
  only over a finite range $[-\Delta,\Delta]$, allowing for a generic
  shape of the host soliton. }
\label{fig}
\end{center}
\end{figure}

Here we will consider a generalized framework where we
expand a set of fields in eigenmodes as \cite{Chibisov:1997rc}
\beq
\Phi^a = \sum_n \mathcal{M}_{n}(\mathbf{e}_\alpha)
  \zeta_n^a(\mathbf{e}_i), 
\eeq
where $\zeta_n$ are eigenfunctions, 
$\mathcal{M}_{n}$ are moduli fields, 
while $\mathbf{e}_\alpha$ and $\mathbf{e}_i$ are 
sets of vectors in transverse (world-volume) dimensions  
($\alpha=0,1,\ldots t$) and codimensions ($i=t+1,\ldots t+c$),
respectively, of a soliton of a generic shape; 
see Fig.~\ref{fig}.
For simplicity we consider only flat space in this Letter and we have
made a decomposition of directions (locally) as
$\mathbb{R}^{d,1}=\mathbb{R}^{t,1}\times\mathbb{R}^c$,
where the $d=c+t$ spatial dimensions are split into $c$ codimensions
and $t$ transverse dimensions.

The kinetic term in the underlying theory will give rise to a kinetic
term for the moduli as
\begin{equation}
\int_{\mathbf{e}_i} |\nabla_\mu\Phi|^2 \supset 
  |\nabla_{\mathbf{e}_\alpha}\mathcal{M}_n|^2
\int_{\mathbf{e}_i} |\zeta_n|^2 \propto \frac{1}{M^c}
|\nabla_{\mathbf{e}_\alpha}\mathcal{M}_n|^2,
\label{eq:kinetic}
\end{equation}
where $M$ is a characteristic mass of the soliton system and $\mu$ are 
all spacetime indices.
For higher-order derivative terms, one similarly obtains e.g., 
\begin{align}
&\int_{\mathbf{e}_i} |\nabla_\mu\Phi|^2 |\nabla_\nu\Phi|^2 \supset
  |\nabla_{\mathbf{e}_\alpha}\mathcal{M}_n|^2
\int_{\mathbf{e}_i} \left|\nabla_{\mathbf{e}_i}\zeta_m\right|^2
  |\zeta_n|^2 \non
&\quad\propto
\frac{1}{M^{c-2}}
|\nabla_{\mathbf{e}_\alpha}\mathcal{M}_n|^2.
\label{eq:enhanced_kinetic}
\end{align}
Notice the relative enhancement of this term compared to that of
\eqref{eq:kinetic}. 
The higher-order term induces an enhanced kinetic term in the
low-energy effective theory living on the soliton.

However, the lower-order term also induces other terms in the
low-energy effective theory, which will be of higher-order.
These induced terms are of a different kind as they are higher-order
corrections coming from integrating out massive modes propagating on
the soliton. 
Let us consider the kinetic term, which would induce something like
\begin{align}
\frac{1}{M^{c+2}}
|\nabla_{\mathbf{\alpha}}\mathcal{M}_n|^2
|\nabla_{\mathbf{\alpha'}}\mathcal{M}_{n'}|^2.
\end{align}
This higher-order correction in the effective theory is naturally
suppressed by (2 powers of) the soliton scale.
Whether this term will be comparable to the higher-order terms in the
theory before we take the low-energy limit on the soliton depends on
the theory and the parameters.

In this Letter, we consider the higher-order terms to be numerically
significant and work in the limit of very high soliton mass, where we
safely can neglect the higher-order corrections coming from
lower-order terms\footnote{Needless to say, this may not always be the
case, but it is a limit we work in here for simplicity. }.

Let us comment on integrating out the host soliton. We assume that
the soliton is extended in the directions spanned by
$\{\mathbf{e}_i\}$ which is taken to be orthogonal to
$\{\mathbf{e}_\alpha\}$. However, integrating over all the subspace
spanned by $\{\mathbf{e}_i\}$ may be problematic; but for physical
reasons we need only integrate over the major energy peak of the
soliton solution (say in the range $[-\Delta,\Delta]$) on which the
moduli live and thus neglect the long tales that the soliton may
possess; see fig.~\ref{fig}. We do this for physically capturing the
low-energy effective theory on the soliton and in a way that we can
still use the decomposition of the transverse and world-volume
coordinates locally.

Finally, we need to assess the quality of the approximation we are
making, since we are taking into account corrections proportional to
powers of the soliton mass coming from higher-order terms. 
The approximation we are making is a separation of scales between the
mass of the host soliton and the energies of the moduli in the
effective action living on the world-volume. 
The higher-order terms, if they have non-negligible coefficients,
induce lower-order terms in the low-energy effective theory on the
soliton which are enhanced by a factor of $(M/m)^{\delta d}$ (where
$\delta d$ is the difference in dimension between the higher-order
term and the lower-order term while $m$ is the typical scale of the
moduli). 
On the other hand, as mentioned above, the lower-order terms also
induce higher-order correction terms which come about from integrating
out massive mod\-es propagating along the soliton. These terms are,
however, suppressed by a factor of $(m/M)^2$ (or higher). 
It has also been assumed all along that the derivatives in the
low-energy effective theory are not too large. 
As long as the ratio $m/M$ is sufficiently small, we can use just the
leading-order low-energy effective theory.

Higher-order corrections coming from the lower-order terms, as
mentioned above, can however be calculated
systematically \cite{Fujimori:2010fk}, but we will not consider them 
in this Letter; i.e., here we present only the leading-order effective 
action.

\section{Non-linear sigma model}

To illustrate our framework more explicitly,
we will now specialize the considerations presented above to
an O(4)-sigma model with higher-derivative terms in $3+1$ flat
dimensions, which has scalar fields, $n^a$, of an O(4) vector, with
$a=1,\ldots,4$ and Lagrangian density
\beq
\mathcal{L} = -m^4 V + c_2 m^2 \mathcal{L}_2 + c_4 \mathcal{L}_4 
+ \frac{c_6}{m^2} \mathcal{L}_6 + \cdots
\label{eq:Lnlsm}
\eeq
where $\mathcal{L}_n$ is the Lagrangian density containing the $n$-th
order derivative terms
\begin{align}
-\mathcal{L}_2 &= \frac{1}{2}\p_\mu\mathbf{n}\cdot\p^\mu\mathbf{n}, 
  \label{eq:L2}\\
-\mathcal{L}_4 &=
\frac{1}{4}\big(\p_\mu\mathbf{n}\cdot\p^\mu\mathbf{n}\big)^2
-\frac{1}{4}\big(\p_\mu\mathbf{n}\cdot\p_\nu\mathbf{n}\big)^2,
  \label{eq:L4}\\
\mathcal{L}_6 &= \mathcal{B}_\mu\mathcal{B}^\mu,\quad
\mathcal{B}^\mu = \frac{1}{6}\epsilon^{\mu\nu\rho\sigma}\epsilon^{a b c d}
\p_\nu n^a\p_\rho n^b\p_\sigma n^c n^d, \label{eq:L6}
\end{align}
and $\mathcal{B}^\mu$ is the baryon current. 
Finally, an appropriate potential should be chosen for the soliton
under study. 
There still remains a choice to be made, i.e.~the codimension of the
soliton under consideration. 
Since we consider $\mathbb{R}^{3,1}$ here, there are only two
non-trivial cases: a codimension-one soliton like a domain wall or a
codimension-two soliton like a vortex. 
We will study each in turn in the following.

\subsection{Codimension-one case}

We will now consider the soliton of the type which is described by a
codimension-one field $\zeta(\mathbf{e}_3)$ and two moduli
$\mathcal{M}_{1,2}(\mathbf{e}_1,\mathbf{e}_2)$, where the
\emph{condensate} field is a function of the direction spanned by the
vector $\mathbf{e}_3$ only and the moduli are functions of two
orthogonal directions $\mathbf{e}_1$ and $\mathbf{e}_2$. 
For concreteness we will parametrize the non-linear sigma-model field, 
$\mathbf{n}$, as 
\beq
\mathbf{n} = \{\mathbf{b}\sin f,\cos f\},
\label{eq:Ansatz}
\eeq
where $\mathbf{b}$ are scalar fields of a unit 3-vector
($\mathbf{b}\cdot\mathbf{b}=1$) describing
two moduli and is a function only of the orthogonal directions to the
field $f$, i.e.~$\mathbf{b}(\mathbf{e}_1,\mathbf{e}_2)$. 
The domain solution also possesses a position modulus, which we will
not take into account in this Letter.
Taking the Lagrangian densities (\ref{eq:L2}-\ref{eq:L6}) one-by-one,
choosing the potential
\beq
m^2 V = -\frac{1}{2} m_3^2 n_3^2 + \frac{1}{2} M^2 (1 - n_4^2),
\eeq
and integrating over the
codimension spanned by $\mathbf{e}_3$, we get
\begin{align}
-\mathcal{L}_2^{\rm eff} &= \frac{a_{2,0}^{\alpha\alpha'}}{M} 
\p_\alpha\mathbf{b}\cdot\p_{\alpha'}\mathbf{b}, \quad
m^2 V^{\rm eff} = -\frac{a_{2,0} m_3^2}{M} b_3^2, \non
-\mathcal{L}_4^{\rm eff} &= a_{2,2}^{\alpha\alpha'} 
  M\,\p_\alpha\mathbf{b}\cdot\p_{\alpha'}\mathbf{b} \non
&\phantom{=}
  + \frac{a_{4,0}^{\alpha\alpha'\beta\beta'}}{2M}
    \big(\p_\alpha\mathbf{b}\times\p_\beta\mathbf{b}\big)\cdot
    \big(\p_{\alpha'}\mathbf{b}\times\p_{\beta'}\mathbf{b}\big), \non
-\mathcal{L}_6^{\rm eff} &= 
a_{4,2}^{\alpha\alpha'\beta\beta'} M 
  \big(\p_\alpha\mathbf{b}\times\p_\beta\mathbf{b}\big)\cdot
  \big(\p_{\alpha'}\mathbf{b}\times\p_{\beta'}\mathbf{b}\big), 
\end{align}
where we have defined the dimensionless constants as follows
\begin{align}
&a_{k,\ell}^{\alpha_1{\alpha_1}'\ldots\alpha_n{\alpha_n}'} \equiv \non
&\frac{M}{2} \int_{\mathbf{e}_3} \sqrt{\det h} \; \sin^{k} f
  \left(\frac{\p_{\mathbf{e}_3} f}{M}\right)^{\ell}
  h^{\alpha_1{\alpha_1}'}\cdots h^{\alpha_n{\alpha_n}'}, \label{eq:adef}
\end{align}
where $h^{\alpha\alpha'}$ is the inverse induced metric on the surface
of the host soliton. Note that to leading order which we consider
here, the induced metric is diagonal. 

The least surprising result is the kinetic term giving back the
kinetic term for the moduli with the normalization constant of the
effective Lagrangian being $1/M$. 
The Skyr\-me term gives also back a Skyrme term for the moduli, but in
addition it induces again a kinetic term for the moduli, however
enhanced by a factor of $M^2$.
Finally and perhaps most interestingly, the sixth-order derivative
term induces only the (baby-)Skyrme term for the moduli and nothing
else. 

Putting the pieces together, we have
\begin{align}
&-\mathcal{L}^{\rm eff} = 
  \left(\frac{c_2 a_{2,0}^{\alpha\alpha'} m}{M} + \frac{c_4 a_{2,2}^{\alpha\alpha'} M}{m}\right) m \;
  \p_\alpha\mathbf{b}\cdot\p_{\alpha'}\mathbf{b} \non
&
  +\left(\frac{c_4 a_{4,0}^{\alpha\alpha'\beta\beta'} m}{2 M} +
  \frac{c_6 a_{4,2}^{\alpha\alpha'\beta\beta'} M}{m}\right) \non
&\times
  \frac{1}{m}\big(\p_\alpha\mathbf{b}\times\p_\beta\mathbf{b}\big)\cdot
  \big(\p_{\alpha'}\mathbf{b}\times\p_{\beta'}\mathbf{b}\big)
- \frac{a_{2,0} m_3^2 m^2}{M} b_3^2.
\label{eq:Lcodim1}
\end{align}
There are now three mass scales in the game: the domain wall mass $M$,
the length scale of the sector that has generated the kinetic terms,
$m$, and finally the mass term for the moduli $m_3$. 
Symmetry breaking requires that $m_3\ll M$; this is also needed in
order for the scales of the moduli to be much smaller than that of the
host soliton. 
The total energy of the domain wall is 
$E_{\rm DW} = 2\sqrt{c_2} M m^2 A$, where $A$ is the area of the
domain wall and the thickness of the domain wall is 
$L_{\rm DW} = 1/M$. Hence in order for the moduli to be really
localized, we need $M$ larger than the other scales in the problem, in
particular $M\gg m$ (or more precisely
$\sqrt{c_6}M\gg\sqrt{c_4}m$). 
In this limit, we can neglect the first term in each parenthesis of
\eqref{eq:Lcodim1}.

Let us comment on the higher-order corrections from the lower-order
terms due to integration out of massive modes. The kinetic term will
induce higher-order corrections of order $(m^2/M^3)$, which in our
regime of parameters will be small compared to the leading-order terms
that we have given here, which are of orders $1/M$ and $M/m^2$,
respectively. 
As long as the ratio $m/M$ is sufficiently small, our leading-order
low-energy effective theory on the soliton is a good approximation.

Let us further comment on leaving out the position moduli from the
low-energy effective action. The position moduli are first of all not
as interesting as the orientational moduli, in the physical context we
have in mind here. Second, once the host soliton is curved, the
position moduli acquire a mass of the order of the curvature scale and
are thus subleading with respect to the orientational part. 
Let us however remark that the higher-order corrections coming from
the kinetic term generically induce a mixing term between the
position and orientation moduli at the fourth order in derivatives. 
These higher-order corrections can, however, be systematically
calculated using the approach of Ref.~\cite{Fujimori:2010fk}.

The domain wall can host a so-called baby-Skyrmion
\cite{Piette:1994ug}, which can be identified with a Skyrmion 
\cite{Skyrme:1961vq} in the bulk
\cite{Nitta:2012wi,Gudnason:2014nba} 
(lower dimensional analogues of this correspondence 
can be found in Ref.~\cite{Nitta:2012xq}.)

Using a scaling argument \cite{Derrick:1964ww}, we can estimate the
size of the baby-Skyrmion on the flat domain wall as
$1/L\sim\sqrt{m_3/M}m$, and as $M\gg m_3$ it is always relatively
large in units of $1/m$ (this estimate holds also for vanishing
$c_4$). 

If $c_6$ is very small or vanishes, the size estimate of the
baby-Skyrmion on the flat domain wall in the large-$M$ limit becomes
$1/L\sim\sqrt{m_3m}$, and so is independent of the thickness of the
host domain wall.

\subsubsection{Example 1: flat domain wall}

The sine-Gordon kink is an exact solution to the equations of motion
derived from the Lagrangian density \eqref{eq:Lnlsm} with the field
parametrization \eqref{eq:Ansatz}
\beq
f=2\arctan\exp(\pm Mx/\sqrt{c_2}).
\eeq
This solution is exact when the moduli, $\mathbf{b}$ are (any)
constants. Due to separation of scales, we can still use this soliton
shape as a good approximation even when the moduli do possess
dynamics. 
The full solutions have been obtained in Ref.~\cite{Gudnason:2014nba}.

A vast simplification in this example is that the induced metric is
just the flat metric, so the effective Lagrangian coefficients, $a$,
do not depend on the Greek indices. 
We can thus evaluate the coefficients in the effective Lagrangian
\begin{align}
a_{k,\ell} &= \frac{1}{2}c_2^{(1-\ell)/2} 
\int_{-\Delta M/\sqrt{c_2}}^{\Delta M/\sqrt{c_2}} dy \; 
\sech^{k+\ell} y.
\end{align}
If we define, $a_{k+\ell}\equiv a_{k,\ell} \, c_2^{(\ell-1)/2}$, we
have 
\begin{align}
a_{k+\ell} &= \frac{1}{2}\sinh y
\left.{}_2F_1\left(\frac{1}{2},\frac{1+k+\ell}{2},\frac{3}{2},-\sinh^2
y\right) \right|_{-\frac{\Delta M}{\sqrt{c_2}}}^{\frac{\Delta
M}{\sqrt{c_2}}}, \nonumber
\end{align}
where ${}_2F_1$ is a hypergeometric function. 
If we take the limit of $\Delta M\to\infty$, the constants are 
$a_{2}=1$, $a_{4}=2/3$, and $a_{6}=8/15$.
$a_{2n}=\sqrt{\pi} \,\Gamma(n)/\Gamma(n+1/2)$, with
$n\in\mathbb{Z}_{>0}$ a positive integer, takes value in $(0,1]$ and
is monotonically decreasing with $n$. 
For the flat domain wall case, we can finally write down the effective
Lagrangian density
\begin{align}
-\mathcal{L}^{\rm eff} &= 
\left(\frac{c_2^{\frac{3}{2}} a_2 m}{M} + \frac{c_4 a_4 M}{\sqrt{c_2}
  m}\right) m \big(\p_\alpha\mathbf{b}\big)^2 \non
&
+ \left(\frac{\sqrt{c_2} c_4 a_4 m}{2M} 
  + \frac{c_6 a_6 M}{\sqrt{c_2} m}\right) \frac{1}{m} 
  \big(\p_\alpha\mathbf{b}\times\p_\beta\mathbf{b}\big)^2 \non
&
- \frac{\sqrt{c_2} a_2 m^2 m_3^2}{M} b_3^2,
\label{eq:Lcodim1_flatDW}
\end{align}
where the world-volume directions are
$\{\mathbf{e}_{\alpha}\}=\{y,z\}$.

\subsubsection{Example 2: spherical domain wall}

This is the first non-flat example. 
Let us begin with a word of caution. Our construction focuses on the
effective description of the moduli on a soliton of a given shape. It
does not guaranty stability or even existence of the host
soliton. These questions are a separate issue and should be addressed
carefully and independently. 
A further issue is that not all topological sectors of the moduli are
available in the effective theory. 
In this case of a spherical domain wall; the moduli have to live in
the topological charge-one sector. 
Full solutions have been constructed in the literature
\cite{Gudnason:2013qba}.  

Here we will simply assume the form of the spherical domain wall with
size $R$ and construct the effective theory that would live on such an
object. $R$ will however be a function of the parameters in the theory
as it is actually determined dynamically.
The induced inverse metric is
$\{h^{rr},h^{\theta\theta},h^{\phi\phi}\}=\{1,R^2/r^2,R^2/r^2\sin^2\theta\}$,
and $\sqrt{\det h}=(r/R)^2\sin\theta$, where we have rescaled the
induced metric so that it is dimensionless.  
The constants of the effective Lagrangian can be determined from
\eqref{eq:adef}. 
Finally, we can write the effective Lagrangian in this case
\begin{align}
-&\mathcal{L}^{\rm eff} = 
m\sin\theta\left[c_2 m R\tilde{a}_{2,0,0} + \frac{c_4}{m
    R}\tilde{a}_{2,2,0}\right] \non
&\phantom{=+}
  \times\left((\p_\theta\mathbf{b})^2 +
  \frac{1}{\sin^2\theta}(\p_\phi\mathbf{b})^2\right) \non
&\phantom{=}
+\frac{m}{\sin\theta}\left[\frac{c_4}{m R}\tilde{a}_{4,0,-2} +
  \frac{2c_6}{(m R)^3}\tilde{a}_{4,2,-2}\right] \non
&\phantom{=+}
  \times
  \big(\p_\theta\mathbf{b}\times\p_\phi\mathbf{b}\big)^2 
-(m R)^2 R m_3^2 \sin\theta\tilde{a}_{2,0,2} b_3^2, 
\end{align}
where the integration measure now is simply $d\theta d\phi$ and the
following dimensionless constants have been defined
\beq
\tilde{a}_{k,\ell,n} \equiv \frac{1}{2}\int dy\; y^{n} \sin^k f(y)
(\p_y f)^\ell, \quad y\equiv \frac{r}{R}.
\eeq
In this example we will not contemplate taking any limits, as the
system is somewhat complicated. The size of the sphere, $R$, is
dynamically determined and is a function of the other parameters in
the theory, e.g.~$c_2,c_4,c_6,m,M$ and so on.

\subsection{Codimension-two case}

The final type of soliton we will consider is of codimension two and
is described by two fields, $f(\mathbf{e}_2,\mathbf{e}_3)$,
$g(\mathbf{e}_2,\mathbf{e}_3)$ and a single modulus
$\mathcal{M}(\mathbf{e}_1)$, where the modulus lives in a single 
dimension only (plus time). We will parametrize the non-linear
sigma-model field, $\mathbf{n}$, as
\beq
\mathbf{n} = \{\sin f\cos g,\sin f\sin g,\mathbf{b}\cos f\},
\eeq
where $\mathbf{b}$ are scalar fields of a unit 2-vector
($\mathbf{b}\cdot\mathbf{b}=1$) describing an
$S^1$ modulus and is a function orthogonal to both the fields of the
host soliton $f,g$, i.e.~$\mathbf{b}(\mathbf{e}_1)$. 
We take again the Lagrangian densities (\ref{eq:L2}-\ref{eq:L6}),
choose the potential
\beq
m^2 V = -\frac{1}{2}m_3^2 n_3^{p_3} + m^2 V_{\rm vortex},
\eeq
where $p_3=1,2$ and integrate over the codimensions $\mathbf{e}_2$ and
$\mathbf{e}_3$ to obtain 
\begin{align}
-\mathcal{L}_2^{\rm eff} &= \frac{a_{2,0,0}^{\alpha\alpha'}}{M^2}
  \p_{\alpha}\mathbf{b}\cdot\p_{\alpha'}\mathbf{b}, 
  \label{eq:Lcodim2L2}\\
-\mathcal{L}_4^{\rm eff} &= a_{2,1,0}^{\alpha\alpha'}\; 
  \p_{\alpha}\mathbf{b}\cdot\p_{\alpha'}\mathbf{b}, 
  \label{eq:Lcodim2L4}\\
-\mathcal{L}_6^{\rm eff} &= 2a_{2,0,2}^{\alpha\alpha'}M^2\; 
  \p_{\alpha}\mathbf{b}\cdot\p_{\alpha'}\mathbf{b}, 
  \label{eq:Lcodim2L6}\\
m^2 V^{\rm eff} &= -\frac{a_{p_3,0,0}m_3^2}{M^2} b_1^{p_3}, 
\end{align}
which are all Lagrangians for a free (massive) theory for the
modulus. The non-trivial part however is the content of the
coefficients
\begin{align}
&a_{p,k,\ell}^{\alpha_1{\alpha_1}'\ldots\alpha_n{\alpha_n}'}
\equiv \frac{M^{2-2k-2\ell}}{2} \non
&
\times \int_{\mathbf{e}_2,\mathbf{e}_3} \sqrt{\det h} \; \cos^p f
\left[(\nabla_{\mathbf{e}_i}f)^2 + \sin^2 f (\nabla_{\mathbf{e}_i} g)^2\right]^k \non
&\phantom{\times\int_{\mathbf{e}_2,\mathbf{e}_3}\ } 
\times
\left[\sin f\epsilon^{ij}\nabla_{\mathbf{e}_i} f \nabla_{\mathbf{e}_j} g\right]^\ell
h^{\alpha_1{\alpha_1}'} \cdots h^{\alpha_n{\alpha_n}'}.
\end{align}
Let us first put together the pieces of the effective Lagrangian
density
\begin{align}
-\mathcal{L}^{\rm eff} &= \left(
  \frac{c_2 a_{2,0,0}^{\alpha\alpha'} m^2}{M^2}
  + c_4 a_{2,1,0}^{\alpha\alpha'}
  + \frac{2 c_6 a_{2,0,2}^{\alpha\alpha'} M^2}{m^2}\right) \non
&\phantom{=+\ }\times
  \p_\alpha\mathbf{b}\cdot\p_{\alpha'}\mathbf{b} 
- \frac{a_{p_3,0,0} m^2 m_3^2}{M^2} b_1^{p_3}.
\end{align}
The first term \eqref{eq:Lcodim2L2} gives just a kinetic term for the
modulus with a standard prefactor, while the second term
\eqref{eq:Lcodim2L4} gives the kinetic term but with a relative
enhancement by a factor proportional to the kinetic energy of the host
soliton. Finally, and perhaps most interestingly, the last term
\eqref{eq:Lcodim2L6} gives again a kinetic term for the modulus, but
with a prefactor proportional to the baby-Skyrmion charge of the
orthogonal $S^2$ to the $S^1$ where the modulus lives. For some host
solitons, this charge may vanish of course.

Let us again comment on the higher-order corrections from the
lower-order terms due to integration out of massive modes. 
The first correction will be a fourth-order term and thus it will not
compete with any terms given here.

\subsubsection{Example 3: straight vortex}

In the straight vortex case, we consider a non-trivial soliton in the
fields $f,g$ and the $S^1$ modulus extending in the $z$ direction.
For this we can choose the potential
\beq
m^2 V_{\rm vortex} = \frac{1}{2} M^2 (n_1^2 + n_2^2)(n_3^2 + n_4^2),
\eeq
and the coefficients can easily be evaluated. For the straight vortex
the only metric-component entering the coefficient is $h^{zz}=1$ which
is trivial and the integrals can be carried out numerically in
cylindrical coordinates using $\det h=\rho^2$ and the polar
derivatives, i.e.~$(\nabla f)^2=f_\rho^2+f_\theta^2/\rho^2$.
The effective Lagrangian density thus reads
\begin{align}
-\mathcal{L}^{\rm eff} &= \left( 
  \frac{c_2 a_{2,0,0} m^2}{M^2}
  + c_4 a_{2,1,0}
  + \frac{2 c_6 a_{2,0,2} M^2}{m^2}\right)
  (\p_z\mathbf{b})^2 \non
&\phantom{=}
- \frac{a_{p_3,0,0} m^2 m_3^2}{M^2} b_1^{p_3}.
\end{align}
This effective theory possesses sine-Gordon kinks. For instance, for
$p_3=1$ a baby-Skyrmion string extended in the $z$ direction can bear
sine-Gordon kinks in terms of its twisted $S^1$ modulus and each of
these kinks can be interpreted in the full 3-dimensional theory as
Skyrmions. 
Setting $p_3=2$ results in half-kinks which correspond in the full
3-dimensional theory to half-Skyrmions (possessing half the baryon
charge of that of a Skyrmion).

The final example is to compactify the vortex on a circle to obtain a 
vorton \cite{Davis:1988jq,Garaud:2013iba}. In this case, the effective
Lagrangian is basically the same as the above one, with the kinetic
term replaced by $(\p_\theta\mathbf{b})^2$ and of course different,
but straightforward to compute, coefficients $a$. In this case, the
twisting of the modulus implies that only full Skyrmions exist (and
thus no half-Skyrmions).

\section{Discussion}

We have developed a basic framework for calculating effective
actions on, in principle, generic solitons. The advantage of this
approach is the simplified theories for the types of objects a host
soliton can host. The disadvantage is that stability and existence
should be carefully examined separately. 
In this work we did not consider the translational type of moduli,
which is related with the shape itself of the soliton as this is a
somewhat more complicated problem; we thus leave this part for future
developments.
Needless to say, one can calculate higher-order classical corrections
to these effective actions and finally one may consider also quantum
corrections etc. 
A different generalization that would be interesting to consider for
GUT-scale solitons is to take the curvature of spacetime into account;
this can be seen as the next-level geometric backreaction.

\subsection*{Acknowledgments}

The work of MN is supported in part by Grant-in-Aid for Scientific
Research (No. 25400268) and by the ``Topological Quantum Phenomena'' 
Grant-in-Aid for Scientific Research on Innovative Areas (No. 25103720)  
from the Ministry of Education, Culture, Sports, Science and Technology 
(MEXT) of Japan. 
SBG thanks Keio University for hospitality where this work was carried
out. 
SBG also thanks the Recruitment Program of High-end Foreign Experts for 
support.


\end{document}